\documentclass[12pt,epsfig]{article}
\usepackage{etex}
\usepackage{epsfig}
\usepackage{epsf,latexsym}
\usepackage{amsfonts,amssymb,amsmath} 
\usepackage{mathtools}
\usepackage{tikz}
\epsfverbosetrue
\newcommand{\double}[1]{\mathbb{#1}}

\newcommand{\rr}{\double{R}}

\newcommand{\llll}{\mathcal{L}}

\newcommand{\dee}{\hbox{\rm{D}}}
\newcommand{\de}{\hbox{\rm{d}}}

\newcommand{\pa}{\partial}

\newcommand{\dpp}{\vcentcolon}
\newcommand{\bb}{\begin{eqnarray}}
\newcommand{\ee}{\end{eqnarray}}
\newcommand{\eee}{\nonumber\end{eqnarray}}
\newcommand{\qq}{\quad}

\begin{document}

\font\twelve=cmbx10 at 13pt
\font\eightrm=cmr8

\thispagestyle{empty}

\begin{center}
${}$
\vspace{3cm}

{\Large\textbf{Torsion, an alternative to dark matter?}} \\

\vspace{2cm}

{\large
Andr\'e Tilquin\footnote{tilquin@cppm.in2p3.fr }
(CPPM\footnote{Centre de Physique des Particules de
Marseille\\\indent${}$\qq\qq CNRS--Luminy, Case
907\\\indent${}$\qq\qq F-13288 Marseille Cedex 9\\\indent${}$\qq
Unit\'e Mixte de Recherche (UMR 6550)
du CNRS et de l'Universit\'e Aix--Marseille 2}),
Thomas Sch\"ucker\footnote{at Universit\'e de Provence,
thomas.schucker@gmail.com } (CPT\footnote{Centre de Physique
Th\'eorique\\\indent${}$\qq\qq CNRS--Luminy, Case
907\\\indent${}$\qq\qq F-13288 Marseille Cedex 9\\\indent${}$\qq
Unit\'e Mixte de Recherche (UMR 6207)
du CNRS et des Universit\'es Aix--Marseille 1 et 2\\
\indent${}$\qq et Sud
Toulon--Var, Laboratoire affili\'e \`a la FRUMAM (FR 2291)})}

\vspace{3cm}

{\large\textbf{Abstract}}
\end{center}
We confront Einstein-Cartan's theory with the Hubble diagram. An affirmative
answer to the question in the title is compatible with today's 
supernovae data.

\vspace{3.6cm}

\vskip 1truecm

\noindent CPT-2011/P004\\
\noindent 1104.0160

\section{Introduction}

The present day standard model of cosmology with a cosmolological constant and pressure-less dark matter  fits oberservational data. However, despite quite some effort, no particle candidate for dark matter has been discovered. On the other hand, the standard model has only few
 challenging models with a finite number of parameters. We propose to use an Einstein-Cartan's theory  as an alternative. The cosmological model we will discuss has the same number of parameters as the standard model. The dark matter density today of the standard model is replaced by a mean spin density of baryonic matter today.
 
 \bigskip
 Einstein-Cartan's theory \cite{car} has several popular motivations:
\begin{itemize}
\item Space-time torsion, that Einstein puts to zero from the start, is allowed without however modifying the Einstein-Hilbert action. Consequently energy-momentum is still the source of space-time curvature with Newton's constant being the coupling constant and the source of torsion is half-integer spin with the same coupling constant.
\item
To describe matter with half-integer spin, one must use an orthonormal frame and the spin connection $\omega $. Consequently the gauge invariance of the Einstein-Hilbert action is manifest with the gauge group being the Lorentz group. Furthermore, Einstein-Cartan's theory
treats this connection as an independent field variable besides the metric. Both features look like  promising steps towards unification of gravity with the standard model of particle physics, which is a gauge theory with a connection as independent field variable.
\item
Torsion closes the diagram of Figure 1.
Let us explain the two arrows labeled `geometry'. One possibility to define and measure curvature is by parallel transporting a vector around an infinitesimal geodesic parallelogram. The Riemann tensor evaluated on the two vectors defining the parallelogram is precisely the infinitesimal rotation mapping the transported vector after the round trip onto the initial vector. Torsion can be defined and measured in a similar fashion. In presence of non-vanishing torsion, the infinitesimal geodesic parallelogram does not close. The translation from the final to the initial point of the `parallelogram' defines a tangent vector, which is precisely the torsion tensor evaluated on the two vectors defining the `parallelogram'.

\newcommand{\ray} {170 pt}
\newcommand{\sep} {15 pt }
\newcommand{\thick} { 0.6pt }

\begin{figure}[h]

\begin{tikzpicture}
\node[outer sep=\sep](r) at (0:\ray) {rotations};
\node[outer sep=\sep](c) at(60:\ray) {curvature};
\node[outer sep=\sep](e) at (120:\ray) {energy- \\ momentum};
\node[outer sep=\sep](tr) at (180:\ray) {translations};
\node[outer sep=\sep](to) at (240:\ray) {torsion};
\node[outer sep=\sep](s) at (300:\ray) {spin};
\draw[<-,line width=\thick]  (r)--(c) 
node[midway,xshift=44pt]
 {geometry};
\draw[<-,line width=\thick]  (c)--(e) 
node[midway,yshift=22pt]
 {Einstein eq.};
\draw[<-,line width=\thick]  (e)--(tr)
node[midway,xshift=-44pt]
 {Noether thm.};
\draw[<-,line width=\thick]  (tr)--(to)
node[midway,xshift=-44pt]
 {geometry};
\draw[<-,line width=\thick]  (to)--(s)
node[midway,yshift=-22pt]
 {Cartan eq.};
\draw[<-,line width=\thick]  (s)--(r)
node[midway,xshift=49pt]
 {Noether thm.};

\end{tikzpicture}
\caption{ Einstein and Cartan's equations, geometric definition of  curvature and torsion, Noether's theorem}
\end{figure}
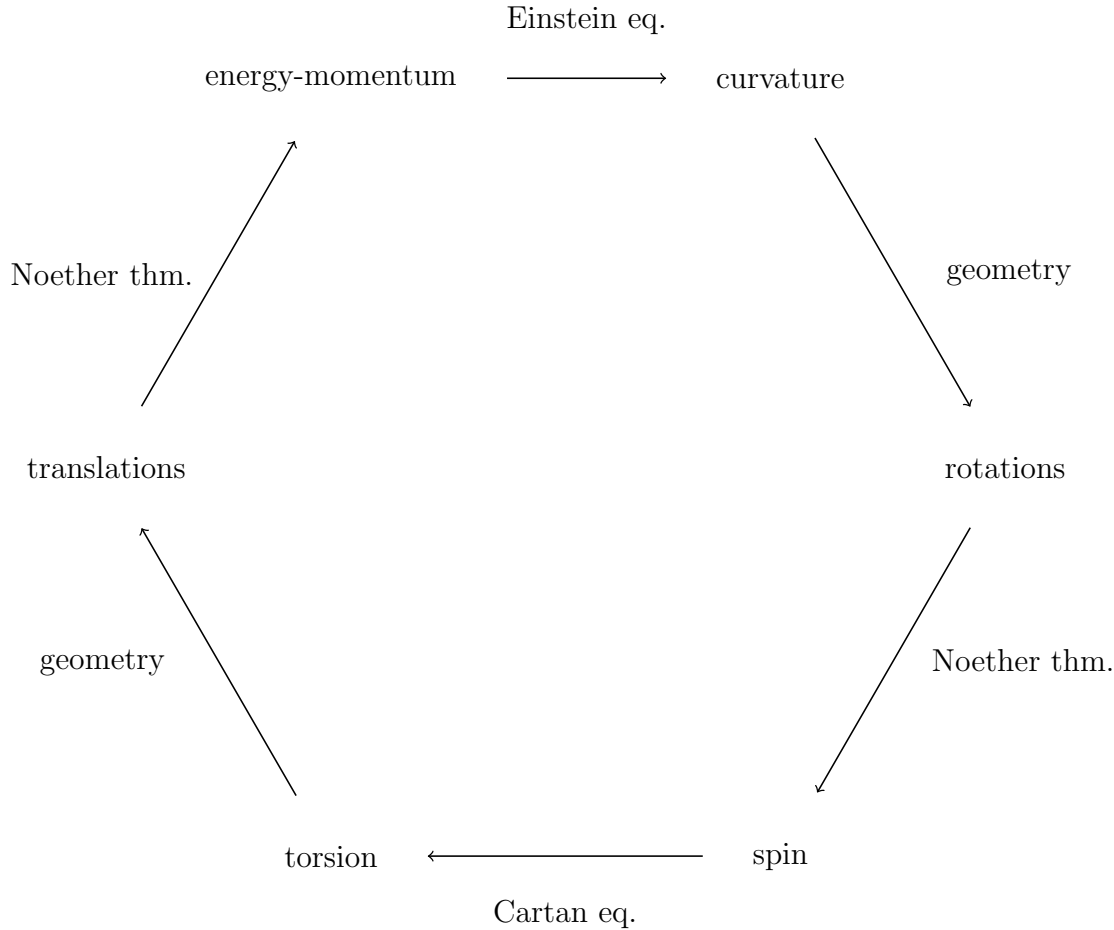

\end{itemize} 

Since a long time physicists have been attracted by the beauty of torsion and there is a vast body of literature on the subject including applications in cosmology. Let us cite {\it the} historical review by Hehl, von der Heyde, Kerlick \& Nester \cite{hehl76} (which was the second author's first contact with torsion), two recent 
reviews \cite{tworecent} and a few contributions of dynamical torsion in cosmology \cite{gmh,dyn}. Other results directly related to the present work with its non-propagating torsion will be referred to as we go along.

Let us mention that supergravity, which has been remarkably popular, is an Einstein-Cartan theory with matter being a spin ${\textstyle\frac{3}{2}} $
Rarita-Schwinger field. If there torsion is put to zero, the theory loses its supersymmetry.

Let us also mention two recent results concerning torsion in noncommutative geometry: \\
{\it (i)} Torsion is produced together with curvature when the flat Dirac operator fluctuates \cite{mm} in the sense of Connes \cite{grav}. \\
{\it (ii)} Connes' spectral action \cite{grav}   produces torsion terms together with the Einstein-Hilbert action \cite{tors}.

\section{Einstein-Cartan's theory in a nut shell}

It is convenient to write Einstein-Cartan's theory using an orthonormal frame, {\it un rep\`ere mobile} using Cartan's words. Let $x^\mu $ ($\mu = 0,1,2,3$) be a coordinate system on an open subset of $\rr^4$ and let $e^a=\dpp {e^a}_\mu \,\de x^\mu,\ a= 0,1,2,3 $ be four 1-forms which are orthonormal with respect to a given space-time metric  $g$. Following the conventions of reference \cite{gs}, we write a metric connection with respect to this orthonormal frame as a 1-form with values in the Lie algebra of the Lorentz group, ${\omega ^a}_b=\dpp{\omega ^a}_{b \mu }\,\de x^\mu $. Then (suppressing all wedge symbols) Cartan's two structure equations read:
\bb R\dpp=\de \omega +{\textstyle\frac{1}{2}} [\omega ,\omega ], \label{cartan1}\ee
for the curvature, a 2-form with values in Lie algebra of the Lorentz group, ${R ^a}_b=\dpp{\textstyle\frac{1}{2}} {R ^a}_{b cd }e^ce^d$, and
\bb T\dpp=\dee e=\de e+\omega e,\label{cartan2}\ee
for the torsion,  a vector-valued 2-form, ${T ^a}=\dpp{\textstyle\frac{1}{2}} {T ^a}_{ bc }e^be^c$.
It will be useful to decompose the torsion tensor  into its three irreducible parts: 
\bb T_{abc}=A_{abc}+\eta_{ab}V_c-\eta_{ac}V_b+M_{abc},\ee
with the completely antisymmetric part $A_{abc}\dpp={\textstyle\frac{1}{3}}(T_{abc}+T_{cab}+T_{bca})$, the vector part $V_c\dpp={\textstyle\frac{1}{3}} T_{abc}\eta^{ab}$, and the mixed part $M_{abc}$ characterized by $M_{abc}=-M_{acb}$, $M_{abc}\eta^{ab}=0$, and $M_{abc}+M_{cab}+M_{bca}=0$.

 We have the two Bianchi identities:
\bb &&\dee R=\de R+[\omega ,R]=0, \\
&& \dee T= \dee\dee e=Re.\ee
In these notations, the Einstein-Hilbert action reads
\bb S_{\rm EH}[e,\omega ]&=&\,\frac{1}{16\pi G}\, 
\int {R^a}_b\ast (e^be_a)-\,\frac{2\Lambda }{16\pi G}\,\int\ast 1\nonumber\\
&=&
\,\frac{-1}{32\pi G}\, 
\int (R^{ab}+{\textstyle\frac{1}{6}}\Lambda e^ae^b )\, e^ce^d\epsilon_{abcd}\nonumber\\
&=&
\,\frac{-1}{16\pi G}\, 
\int ({R^{ab}}_{ab}+2\Lambda )\,\de V,
  \ee
where $\ast$ is the Hodge star of the metric $g$ and $\epsilon_{0123}=1$. 
The energy-momentum current is the vector-valued 3-form $\tau_a$ obtained by varying the orthonormal frame in the matter Lagrangian:
\bb \llll_{\rm M}[e+f,\omega ]-\llll_{\rm M}[e,\omega ]=\dpp -f^a\tau_a+O(f^2).\ee
The energy-momentum tensor $\tau_{ab}$ is defined by $\ast \tau_a=\dpp \tau_{ab}e^b$.

Likewise, the spin current is the Lorentz-valued 3-form $S_{ab}$ obtained by varying the connection in the matter Lagrangian:
\bb \llll_{\rm M}[e,\omega+\chi  ]-\llll_{\rm M}[e,\omega ]=\dpp -{\textstyle\frac{1}{2}} \chi ^{ab}S_{ab}+O(\chi ^2).\ee
The spin tensor $S_{abc}$ is defined by $\ast S_{ab}=\dpp S_{abc}e^c$.

Now we can derive the Einstein equation by varying the total action with respect to the orthonormal frame:
\bb
(R^{ab}+{\textstyle\frac{1}{3}}\Lambda e^ae^b)\,e^d\epsilon_{abcd} =-16\pi G\tau_c \qq {\rm or\  equivalently}\qq
 G_{ab} -\Lambda \eta_{ab} =8\pi G\tau_{ba},\ee
with  the Einstein tensor $G_{ab}\dpp={R^c}_{acb}-{\textstyle\frac{1}{2}} {R^{cd}}_{cd}\,\eta_{ab}$. 

Likewise Cartan's equation is derived by varying the total action with respect to the connection:
\bb T^ce^d\epsilon_{abcd}=-8\pi GS_{ab},
\ee
or equivalently:
\bb A_{cab}+2V_a\eta_{bc}-2V_b\eta_{ac}+M_{cab}=-8\pi GS_{abc}.\ee
Unlike curvature, torsion does not propagate: it is non-vanishing only inside matter with half-integer spin.

Note that in presence of general torsion, the Einstein tensor $G_{ab}$ and
  energy-momentum tensor $\tau_{ab}$ are neither  covariantly conserved nor symmetric. Indeed combining Einstein's equation with the first Bianchi identity, we have:
  \bb (R^{ab}+\Lambda e^ae^b)\,T^d\epsilon_{abcd}=-16\pi G(\dee\tau)_c,\ee
  and combining Cartan's equation with the second Bianchi identity, we have:
  \bb \tau_{ab}-\tau_{ba}=\ast(\dee S)_{ab}.\ee
   
In absence of half-integer spin, general relativity is usually written with respect to a holonomic frame 
$\de x^\mu $ and the metric $g$ is given by its metric tensor $g_{\mu \nu}(x)$ with respect to the coordinates $x^\mu $. The holonomic frame will be useful below to compute geodesics and invariant connections.  The metric tensor reads $g_{\mu \nu}(x) = {e^a}_\mu(x)\,{e^b}_\nu(x)\,\eta_{ab}$. Traditionally the components of the connection with respect to a holonomic frame are denoted by ${\Gamma^\alpha  }_\beta  =\dpp {\Gamma^\alpha  }_{\beta \mu }\,\de x^\mu $, a $g\ell(4)$ valued 1-form. The link between the components of the same connection with respect to the holonomic frame $\Gamma $ and with respect to the orthonormal frame $\omega $ is given by the $GL(4)$ gauge transformation with $e(x)={e^a}_\mu(x)\,\in GL(4)$;
\bb \omega =e\Gamma e^{-1}+e\de e^{-1},
\ee
or with indices:
\bb {\omega ^a}_{b \mu }={e^a}_\alpha\,  {\Gamma^\alpha  }_{\beta \mu }\,{e^{-1\,\beta }}_b +{e^a}_\alpha\,\frac{\pa}{\pa x^\mu }\,   {e^{-1\,\alpha  }}_b.\label{gaugetrans}\ee
The given connection (with or without torsion) is metric if `angles' and `lengths' are preserved under parallel transport. For the components $\omega $ of the connection this just means that its values ${\omega^a}_b $ are in the Lie algebra of the Lorentz group: $\omega _{ab}=-\omega _{ba}$. In terms of the holonomic components the metricity of the connection reads:
\bb \,\frac{\pa}{\pa x^\lambda }\, g_{\mu \nu}
-{\Gamma ^{\bar\mu  }}_{\mu \lambda } g_{\bar\mu \nu}
-{\Gamma ^{\bar\nu }}_{\nu \lambda } g_{\mu\bar \nu}=0.\label{metric}\ee

In presence of general torsion, there are two different geodesics $x^\mu (p)$. The first is defined by parallel transport of the velocity vector $\dot x$ with the given connection:
\bb \ddot x^\lambda +{\Gamma^\lambda   }_{ \mu\nu  }\,\dot x^\mu \,\dot x^\nu=0,\ee
where the over-dot denotes the derivative with respect to the affine parameter $p$. The second geodesic is defined by using the Christoffel connection of the given metric instead of the independent (metric) connection. This second geodesic minimizes locally the `arc length'. Both geodesics coincide if the torsion only has a completely antisymmetric part, $V=0,\ M=0$.

\section{Vector fields preserving metric and connection}

Let $\varphi $ be a diffeomorphism with Jacobian matrix
\bb {\Lambda ^{\bar \mu }}_\mu(x) \dpp= 
\,\frac{\pa\varphi ^{\bar \mu }\ \ }{\ \ \pa x ^\mu }\, (x).
\ee
If $g_{\mu \nu}(x)$ is the metric tensor of a metric $g$ with respect to the coordinates $x^\mu $, then by definition $\varphi $ is a (local) isometry  if
\bb
g_{\mu \nu}(x) ={\left( \Lambda ^T\right)_\mu }^{\bar\mu}(x)\, g_{\bar\mu \bar\nu} (\varphi (x))\,{\Lambda ^{\bar \nu }}_\nu(x)\,.\label{iso}
\ee
Upon linearisation $\varphi (x)=x+\xi (x)+o(\xi ^2)$, where $\xi=\xi ^\alpha \,\pa/\pa x^\alpha  $ is a vector field, equation (\ref{iso}) becomes the Killing equation:
\bb 
\xi ^\alpha \,\frac{\pa}{\pa x^\alpha }\, g_{\mu \nu}+\,\frac{\pa \xi ^{\bar\mu }}{\pa x^\mu }\, g_{\bar\mu \nu}+\,\frac{\pa \xi ^{\bar\nu }}{\pa x^\nu }\, g_{\mu \bar\nu}=0.\label{kill}\ee
 Likewise if ${\Gamma ^\lambda }_{\mu \nu}(x)$ are the components of a connection (not necessarily the Christoffel connection of the metric) with respect to the coordinates $x^\mu $, then by definition $\varphi $ preserves the connection if
 \bb {\Gamma ^\lambda }_{\mu \nu}(x)=
 {\left(\Lambda ^{-1\,T}\right)_{\bar\lambda }}^\lambda (x)\,
  {\Lambda ^{\bar \mu }}_\mu(x)\, {\Lambda ^{\bar \nu }}_\nu(x)  \,{\Gamma ^{\bar\lambda }}_{\bar\mu \bar\nu}(\varphi (x))-
  {\left(\Lambda ^{T}\right)_{\nu }}^{\bar\nu}(x)\,\,\frac{\pa}{\pa x^\mu}\, {\left(\Lambda ^{-1\,T}\right)_{\bar\nu }}^\lambda (x)\,.
\ee
Linearising we obtain:
\bb 
\xi ^\alpha \,\frac{\pa}{\pa x^\alpha }\,  {\Gamma ^\lambda }_{\mu \nu}
-\,\frac{\pa \xi ^{\lambda  }}{\pa x^{\bar\lambda } }\,  {\Gamma ^{\bar\lambda }}_{\mu \nu}
+\,\frac{\pa \xi ^{\bar\mu }}{\pa x^\mu }\,  {\Gamma ^\lambda }_{\bar\mu \nu}
+\,\frac{\pa \xi ^{\bar\nu }}{\pa x^\nu }\,  {\Gamma ^\lambda }_{\mu \bar\nu}
+\,\frac{\pa ^2\xi ^\lambda }{\pa x^\mu \pa x^\nu}\, 
=0.\label{kill2}\ee

\section{The most general Riemann-Cartan space invariant under $O(3)\ltimes\rr^3$}

In order to combine Einstein-Cartan's theory with the cosmological principle we have to solve the Kiling equation (\ref{kill}), its analogue for the connection (\ref{kill2}) and the metricity condition (\ref{metric}) for all vector fields generating the maximal isometry group of space. To simplify the calculations -- and not because we particularly believe  in flat spaces -- we will take this group to be the 3-dimensional Euclidean group $O(3)\ltimes\rr^3$ generated by three infinitesimal rotations $\xi =y\pa/\pa z-z\pa/\pa y,\ z\pa/\pa x-x\pa/\pa z$
and $x\pa/\pa y-y\pa/\pa x$, and by three infinitesimal translations $\xi =\pa/\pa x,\ \pa/\pa y,$ and $\pa/\pa z$, where $x,\,y,\,z$ are Cartesian coordinates.

The most general solution of the Killing equation is well known $\de\tau^2=h\,\de t^2- a^2(\de x^2+\de y^2 +\de z^2)$, with two positive functions $a(t)$ and $h(t)$. By a redefinition of the time coordinate $t$, we set $h\equiv 1$.

Equation (\ref{kill2})  is the analogue of the Killing equation and describes connections invariant under a vector field $\xi $. Technically it  consists of 64 differential equations for each vector field, of which some are empty, some are identical, some are algebraic. Our task therefore is straight-forward and tedious: write down and solve the $6\times 64$ equations (\ref{kill2}) coming from the six vector fields $\xi $. For the three translations we get immediately that all 64 components of $\Gamma $ are independent of $x,\,y$  and $z$. For the three  rotations, it is sufficient to solve equation (\ref{kill2}) for two infinitesimal rotations. It is then automatically satisfied for the third rotation because this rotation is the commutator of the two others. It remains to treat $2\times 64$ equations. Finally, 
the most general connection solving  equation (\ref{kill2}) with the six vector fields has the following non-vanishing components:
\bb {\Gamma ^t}_{t t}=A(t),&
{\Gamma ^t}_{i j}=a(t)b(t)\,\delta _{ij},&
{\Gamma ^i}_{j t}=c(t)\,\delta _{ij},\\
{\Gamma ^i}_{t j}=d(t)\,\delta _{ij},&
{\Gamma ^i}_{jk}=f(t)\,\epsilon _{ijk},&
\epsilon _{123}=1,\ee
with five arbitrary functions of $t$: $A,\,b,\, c,\,d,\,f$.
The metricity condition (\ref{metric}) yields with $'\dpp=\de/\de t$,
\bb A=0,\qq  c=\,\frac{a'}{a},\qq d=\,\frac{b}{a}.\ee 
So far we have only taken into account transformations close to the identity and we still have to discuss space inversion. Since $a$ and $b$ are parity-even and $f$ is parity odd, we must set $f=0$.

This connection reduces to the Christoffel connection if $b=a'$.

We use the gauge transformation (\ref{gaugetrans}) with ${e^a}_\mu = \text{diagonal}(1,a,a,a)$ to compute the non-vanishing holonomic components ${\omega ^a}_{b\mu }$ of the connection:
\bb {\omega ^0}_{ij}={\omega ^i}_{0j}=b\, \delta _{ij}\,.\ee
The Riemann tensor has the following non-vanishing components:
\bb 
{R^0}_{i0k}={R^i}_{00k}=\,\frac{b'}{a}\,\delta _{ik},\qq
{R^i}_{jk\ell}=\frac{b^2}{a^2}\,(\delta _{ik}\delta _{j\ell}- \delta _{i\ell}\delta _{jk}).\ee
The Einstein tensor has:
\bb G_{00}=3\,\frac{b^2}{a^2}\,,\qq
G_{ij}=-\left(2\,\frac{b'}{a}\, +\,\frac{b^2}{a^2}\,\right)\,\delta _{ij}.\ee 
Note that it is symmetric but not covariantly conserved.

The torsion tensor has:
\bb
{T^i}_{0j}=\,\frac{a'-b}{a}\,\delta _{ij}\,.\ee
 In particular the vector part has only a time component, $V_0=(b-a')/a$, and the antisymmetric and mixed parts vanish, $A=0,\,M=0$. This result agrees with the torsion found in references \cite{symt,gmh} for spacetimes with maximally symmetric 3-spaces.
 
 \section{Hubble diagram}
 
 To make contact with physics, we must solve the geodesic equations for co-moving galaxies and for photons \cite{berry}. For both, torsion decouples and they reduce to the geodesic equations with the Christoffel connection of the metric. Consequently the redshift is still given by $z=a_0/a(t)\,-1$ with look-back time $t$ and the apparent luminosity $\ell$ is still related to the absolute luminosity of the standard candle $L$ by
 \bb \ell(t)= \,\frac{L}{4 \pi a^2_0\,x(t)^2}\, \,\frac{a(t)^2}{a^2_0}\, .\label{luminosity} \ee
We have put the earth at the origin of the Cartesian coordinates and the supernova on the $x$-axis:
\bb x(t):=\int_t^{t_0}\,\frac{\de\tilde t}{a(\tilde t)}\, .\ee
It is remarkable, that the Hubble diagram in absence of peculiar velocities is entirely determined by the metric and does not feel torsion. This is not true if the standard candle has non-vanishing peculiar velocity and/or if the messengers are massive neutrinos instead of photons.

On the other hand, the Einstein equations in presence of half-integer spin do feel torsion. Note however that the link between the Hubble constant $H_0$ and $\de (z^2\ell )/\de z (0)$ is purely kinematical. It does not depend on the dynamics, i.e. the Einstein equations. Therefore it does not depend on torsion. This fact will be crucial to identify consistently the initial conditions of the Einstein equations.

\section{Matter, equations of state  and field equations}

Before we can write down the Einstein equations and see how they are modified by torsion, we must talk about their right-hand sides.

The most general energy-momentum tensor invariant under $O(3)\ltimes\rr^3$ must satisfy an equation similar to the Killing equation (\ref{kill}). It therefore contains two arbitrary functions of cosmic time, $\rho (t)$ and $p(t)$ interpreted as energy density and pressure: $\tau_{\mu \nu}= \text{diag}(\rho ,a^2p,a^2p,a^2p)$. Note that we have already used the freedom of coordinate transformations to get rid of one function, $h(t)$, in the metric. We are therefore stuck now with two functions in the energy-momentum tensor. But then we have three functions, $a,\,\rho ,\,p$ and only two independent components in the Einstein tensor and must get rid of one function. This is usually done by an equation of state $p(t)=\dpp w\,\rho (t)$. To simplify even more, one thinks of highly diluted particles and sets the pressure to zero, $w=0$.
 
 Likewise the most general $O(3)\ltimes\rr^3$-invariant spin tensor has one function of time $s(t)$ in the time component of the vector part, $s(t)\dpp = {\textstyle\frac{1}{3}} S_{0ab}\eta^{ab}$. In Einstein-Cartan's cosmology for dust we then have four functions, $a,\, b,\, \rho ,\, s$ but only three equations, two of Einstein and one of Cartan. Therefore we need another equation of state. Our motivation is to do without dark matter. Visible matter consists mostly of protons, neutrons, electrons and neutrinos, all of spin ${\textstyle\frac{1}{2}} $ and all with small densities today. Therefore we keep neglecting the pressure. We also neglect spin-spin interactions and assume the spin density to be proportional to the particle density or energy density: 
 \bb s(t)=\dpp  w_s\rho (t).\ee
  Note that, while $w$ is dimensionless, $w_s$ has units of time. For a single proton, we would expect $w_s\sim \hbar/(m_Pc^2)\sim 10^{-25}$ s. For several protons with some spins  anti-parallel, $w_s$ could be even smaller. For a neutrino, $w_s$ should be larger.
  
We have not tried to derive the spin density from any microscopic model.  Kopczy\'nski \cite{kop}  assumed a Weyssenhoff fluid which has the draw-back of producing a spin-density incompatible with maximal space-symmetry. B\"ohmer \& Burnett \cite{bb} studied Elko spinors or ``dark'' spinors as microscopic model. These spinors are $CPT$-odd and couple to ordinary matter only via gravity and via the Higgs mechanism, motivating the second terminology. Furthermore, they produce a spin-density compatible with the cosmological principle. Pop{\l}awski \cite{pop} uses a particular form for the spin density, which he derives from quarks and leptons.

Now we are in position to write down the generalised Friedmann equations, i.e. the $tt$ and the $xx$ components of Einstein's equations, and Cartan's equation:
\bb 3\,\frac{b^2}{a^2}\, &=&\Lambda +8\pi G\rho ,\label{tt}\\[2mm]
2\,\frac{b'}{a}\, +\,\frac{b^2}{a^2}\, &= &\Lambda ,\label{xx}\\[2mm]
2\,\frac{a'-b}{a}\, &=&8\pi Gw_s\rho \label{cartan}.\ee
We have three equations for three unknown functions, $a,\,b$ and $\rho $. The first equation (\ref{tt}) is algebraic, the other two equations, (\ref{xx}) and (\ref{cartan}), are first order differential equations. We use the algebraic one to eliminate $\rho $ from Cartan's equation (\ref{cartan}). Then we have a unique solution with two inital conditions $a(0)=a_0$ and $b(0)=b_0$. We therefore have four parameters, $a_0,\, b_0,\, \Lambda $ and $w_s$. (We assume Newton's constant known.) These four parameters are not independent. With equation (\ref{tt}) we now eliminate $b_0$, Cartan's equation then relates $a_0,\, \Lambda $ and $w_s$ to the Hubble constant $H_0$. The remaining equation (\ref{xx}) allows us to compute the decceleration parameter $q_0$. We do not spell out this complicated expression for $q_0$ preferring to obtain the Hubble constant from the Hubble diagram at $z=0$ (without using any dynamics) and to obtain the cosmological constant and the state parameter $w_s$ from a fit of the entire Hubble diagram. Let us write the relation between the Hubble constant and $a_0,\, \Lambda $ and $w_s$:
\bb
1=\Omega _{m0}+\Omega _{\Lambda0} +2\Omega _{s0}-\Omega_{s0}^2, \label{friedman} \ee
 using familiar dimensionless quantities:
\bb
\Omega _m\dpp=\,\frac{8\pi G\rho }{3H^2}\,, \qq
\Omega _\Lambda \dpp=\,\frac{\Lambda  }{3H^2}\,, \qq
\Omega _s\dpp= w_s\,\frac{8\pi G\rho }{2H}\,.
\ee
In particular, we see that the scale factor today $a_0$ has dropped out. This is well-known for  cosmology with vanishing spatial curvature  and remains true in presence of non-vanishing torsion.

\section{Data analysis}

To confront Einstein-Cartan's theory with experimental data, we have used the so-called Union 2 sample \cite{union2} containing 577 type 1a Supernovae up to a
redshift of 1.4. Our results have been obtained using the full covariance matrix including correlations and systematic errors.
The magnitude evolution of supernovae with redshift is written as 
$M(z) = m_s - 2.5 \log \ell(z)$ where $m_s$ is a normalisation parameter
fitted to the data and $\ell(z)$ the apparent luminosity (\ref{luminosity}). 
As in standard cosmology, the magnitude evolution with redshift is Hubble-constant free.

The apparent luminosity is computed by solving the differential equations 
(\ref{xx}) and (\ref{cartan}) where the algebraic equation (\ref{tt}) has been used to eliminate $\rho $. 
This system of 2 coupled differential equations is solved numerically by using the Runge-Kutta algorithm \cite{runge} 
with an adapted step in time such that the equivalent redshift step is much smaller than the experimental 
redshift error: $\Delta z < 10^{-5}$.

The best fitted cosmology is obtained using the MINUIT \cite{minuit} package for the iterative $\chi^2$ minimisation with:
\bb \chi^2 = \Delta M^T V^{-1} \Delta M \ee 
where $\Delta M$ is the vector of differences between measured  and expected magnitudes  and $V$ the full covariance matrix.

Since correlations between parameters are expected to be high, we
marginalised over unwanted parameters and compute errors or contours 
by using the frequentist prescription \cite{pdg} and solving 
$\chi^2 = \chi^2_{min} + s^2 $ ($s=1$ for a $1\sigma$ error on a single
variable) where $\chi^2_{min}$ is the minimum $\chi^2$ over all unwanted parameters. 
The Einstein-Cartan cosmology fit is performed with 3 free parameters ($m_s,
\, \Omega_m, \, \Omega_s$)
while $\Omega_\Lambda$ is a derived parameter obtained from the Friedmann-like equation (\ref{friedman}). 
All of the following results are given with marginalisation over the nuisance parameter $m_s$.

Table 1 presents the results for the fit of Einstein-Cartan's theory and for comparison of the pure
Einstein theory. To get the same number of free parameters, 
the fit for the pure Einstein theory does not use the flatness constraint.
The minimum $\chi^2$ for both fits are very close which implies that both
theories are compatible with supernovae data at the same statistical level.
The preferred value for $\Omega_s$ is $0.12$ compatible with zero at a level better than 1 sigma.
The effect of torsion is to lower the preferred value of  $\Omega_m$ to 0.09
which is statistically compatible either with the baryon density of
$0.046 \pm 0.003$  or the total matter density of $0.27 \pm 0.03$  published
by the WMAP collaboration \cite{wmap}. 
However, the spin density can contribute to the dark matter energy density to
a certain amount. On the contrary the cosmological constant is only mildly modified by torsion.

\begin{table}[htbp]
\begin{center}
\label{results}
\begin{tabular}{|c|c|c|c|c|} \hline
                 & $\Omega_m$              & $\Omega_\Lambda$         &       $\Omega_s$  &$\chi^2_{min}$ \\ \hline
Einstein          & $0.35^{+0.24}_{-0.29}$  & $0.88^{+0.42}_{-0.55}$  &       $0.$       &$530.0$  \\ \hline
Einstein-Cartan   & $0.09^{+0.30}_{-0.07}$  & $0.83^{+0.10}_{-0.16}$  &  $0.12^{+0.02}_{-0.22}$ &$530.4$ \\ \hline
\end{tabular}
\caption[]{Fit results (1$\sigma$ errors) for Einstein and Einstein-Cartan theories. No flatness constraint is imposed in Einstein's theory.}
\end{center}
\end{table}

\begin{figure}[h]
\hspace{0.2cm}
\includegraphics[width=16cm, height=16cm]{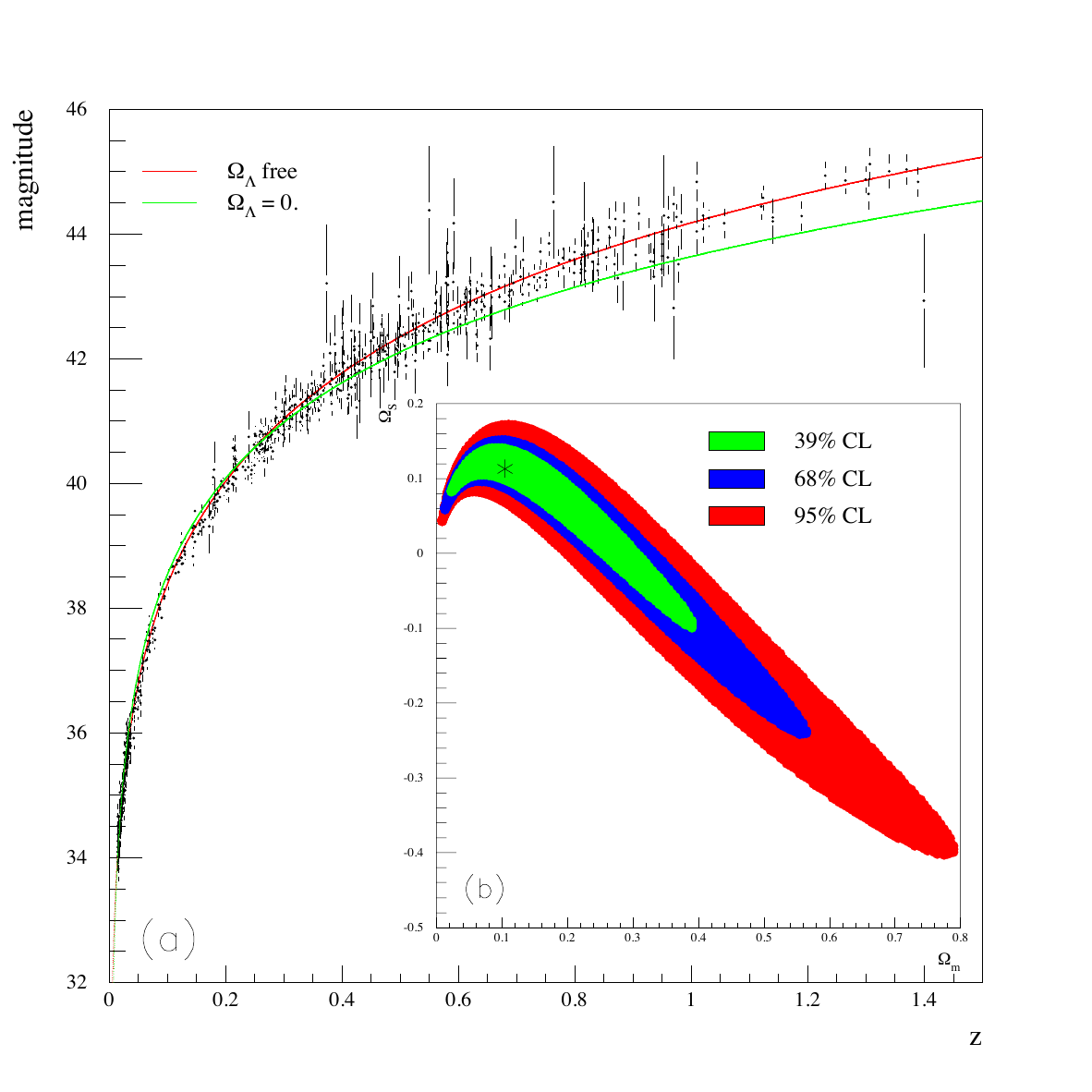}
\caption[]{\small (a) Fit results using the Union 2 Supernovae sample. The
  red (upper) curve corresponds to the Einstein-Cartan 3-fit ($m_s,\Omega_m,\Omega_s$) 
  while the green (lower) curve represents the 2-fit assuming a vanishing
  cosmological constant. (b) $39\%,68\%$ and $95\%$ confidence level contour
  in the ($\Omega_m,\Omega_s$) plane for the Einstein-Cartan 3-fit.}
\label{fig1}
\end{figure}

\begin{figure}[ht]
\begin{center}
\includegraphics[width=16cm, height=16cm]{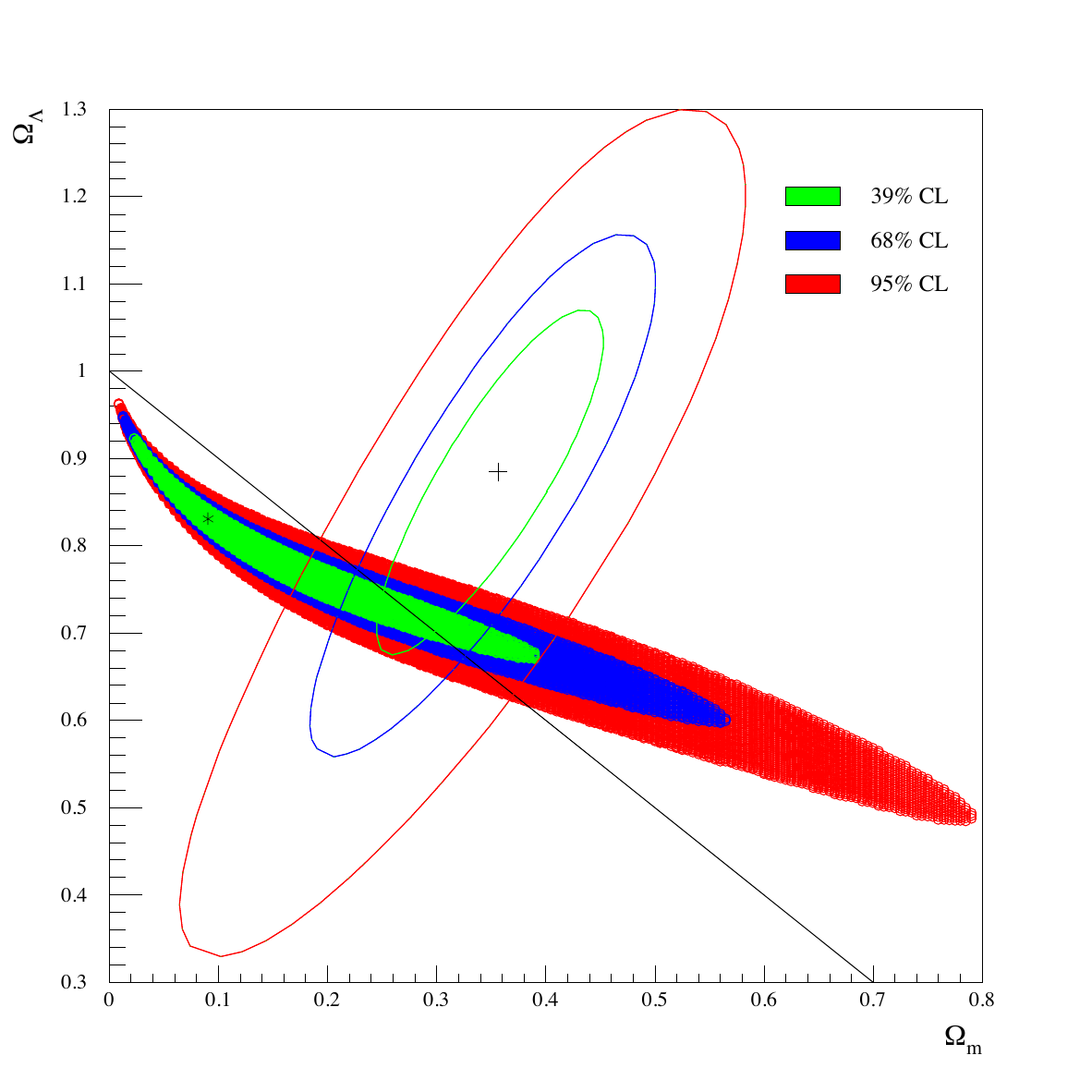}
\caption[]{\small $39\%, 68\%$ and $95\%$ confidence level contours
  in the ($\Omega_m,\Omega_\Lambda$) plane for the Einstein-Cartan 3-fit (full
  contours) and for the Einstein 3-fit (full colored (elliptical) lines). The black (straight) line corresponds to the
  usual Einstein flatness.}
\label{fig2}
\end{center}
\end{figure}

  Figure \ref{fig1}(b) shows probability contours in the ($\Omega_m,\Omega_s$) 
plane. As already mentioned, the degeneracy between both parameters is very
high. As we can see, for negative spin density, the matter 
density increases while the cosmological constant density decreases (Figure \ref{fig2}). This
suggests that a negative spin density may act as dark energy density.

We test this hypothesis by setting $\Omega_\Lambda$ equal to zero in the Einstein-Cartan equations and we redo the fit with only $m_s$ and $\Omega_s$  as free
parameters. In Figure \ref{fig1}(a) the green (lower) curve shows the result of the
best fit for Einstein-Cartan theory fixing $\Omega_\Lambda$ to zero, while 
the red (upper) curve shows the best fit with free cosmological constant energy
density. In the first case, the minimum $\chi^2$ is equal to 1188 for a number
of degrees of freedom of 575 while in the second case the minimum $\chi^2$ is 
530.4 for a number of degrees of freedom of 574. 
These results rule out the hypothesis of torsion to replace dark energy.
Let us mention the analysis of reference \cite{torLam}, which  contains a similar analysis using the parity-odd torsion \cite{bloo}. They find that this part of torsion can mimic a cosmological constant. 

  In Figure \ref{fig2} we show the probability contours in the
 ($\Omega_m,\Omega_\Lambda$) plane for both Einstein-Cartan and Einstein
 theories. The degeneracy between both theories is almost orthogonal. 
The main reason is coming from the Friedmann-like equation for Einstein-Cartan
theory. This constraint mimics an approximate flatness $\Omega_m + \Omega_\Lambda = 1$. 
We notice that both contours cross at a confidence level of $39\%$ at the usual
$\Lambda$CDM  point, corresponding to vanishing torsion. 
From a purely statistical point of view, the actual supernovae data 
do not allow us to discriminate between both theories.   

\section{Conclusion}

It is well known that curvature is not an alternative to dark matter in the Hubble diagram, see Figure 3. On the other hand, within the error bars of the supernova luminosities, torsion \`a la Einstein-Cartan allows us to fit the data with a matter contribution of $\Omega _m=5\, \%$ corresponding to visible matter only. However the best fit has $\Omega _m=9\, \%$ implying still a few percent of dark matter. More disturbingly, its corresponding state parameter is $w_s\sim H_0^{-1}\sim 10^{17}$ s,  42 orders of magnitude away from the naive value
$w_s\sim \hbar/(m_Pc^2)\sim 10^{-25}$ s. 

To conclude, we think that the present result is encouraging enough to reconsider the rotation curves in galaxies and the CMB anlysis in the light of the Einstein-Cartan theory.

\vskip.5 cm \noindent {\bf Acknowledgement:} It is a pleasure to thank Sami ZouZou for his advice.

\end{document}